\documentclass[11pt]{article}
\usepackage{epsf}
\def\demi{{\textstyle {1\over2}}}

\def\L{l}

\def\pa{\partial}
\def\de{\delta}

\def\bea{\begin{eqnarray}}
\def\eea{\end{eqnarray}}

\def\be{\begin{equation}}
\def\ee{\end{equation}}

\def\Q{s_n }

\let\nonu=\nonumber

\begin{document}
\bibliographystyle{perso}

\begin{titlepage}
\null \vskip -0.6cm
\hfill PAR--LPTHE 02--01

%\hfill RUNHETC-2000-54

\hfill hep-th/0201110

\vskip 1.4truecm
\begin{center}
\obeylines

        {\Large	Quantum Field Theory with Extra Dimensions}
\vskip 6mm
Laurent Baulieu$^\dagger$ 
 % and Daniel Zwanziger $^*$
{\em  $^\dagger$  LPTHE,~ Universit\'es Pierre et Marie Curie Paris 6 
et Denis-Diderot~Paris~7}
 %{\em and}
%{\em
% $^{\dag}$  Physics, Rutgers University, New Brunswick,
%NJ~60637,~USA }
%{\em and}
% {\em
% $^*$  Physics Department, New York University,  Washington Place, New York,
%  NY~10021,~USA }

\end{center}

\vskip 13mm

\noindent{\bf Abstract}: We explain that a bulk with arbitrary
dimensions can be added to the space over which a quantum field
theory is defined. This gives a TQFT such that its correlation functions  
in a slice are the same as those of the original quantum field theory. 
This  generalizes the stochastic quantization 
scheme,   where the bulk is one dimensional.  \vfill

%\begin{center}

\hrule \medskip
%\obeylines
{\em \noindent
Postal address: 
Laboratoire de Physique Th\'eorique et des Hautes Energies,
 UMR CNRS 7589, 
 Universit\'e Pierre et Marie Curie,
 bo\^\i te postale 126,
4,~place Jussieu, F--75252 PARIS Cedex 05.}

%\end{center}
\end{titlepage}

\section{Introduction}

  In recent works~\cite{BaZw}\cite{BaGrZw}, the basic ideas of stochastic
quantization~\cite{pawu}\cite{huffel} have been elaborated in a systematic
approach, called bulk quantization.  These papers  give a central role to the
introduction of a symmetry of a topological character.  The correlation
functions for equal values of the bulk time define the correlations of the
physical theory.  Pertubatively, bulk quantization and the usual
quantization method are equivalent because the observables satisfy the same
Schwinger-Dyson equations in both approaches; basic concepts such as the
definition of the $S$-matrix (in the LSZ sense) and the Cutkowski rules can
be also directly addressed in 5 dimensions~\cite{BaZw}. We believe that the
difficulty of giving a consistent stochastic interpretation to all details
of the formalism, especially in the case of gauge theories, justifies to
directly postulate that bulk quantization is a particular type of a
topological field theory. Moreover, there is an interesting geometrical
interpretation for many of the ingredients that are needed in  bulk
quantization. The idea of a topological field theory is actually relevant,
since one wishes to define observable that are independent of most of the
details of the bulk, such as the metrics components $g_{tt}$ and $g_{\mu
t}$.  Quite interestingly, the interpretation of anomalies in gauge
theories is that, the limit of taking the limit of an equal bulk time can be
ambiguous.

As shown in~\cite{BaZw}, the additional dimension $t$ does not take part in
the Poincar\'e group of symmetries for the case of an additional non
compact dimension. The homogeneity of the Lagrangian requires that $t$ has
the dimension of the square of the ordinary coordinates. We investigated
how this method can be applied to gravity and supersymmetric theories in
\cite{gravity}\cite{susy}.  Here, we will show that bulk quantization can
be generalized to the case where the bulk has an arbitrary dimension $n\geq
1$, according to the following picture:
\vskip .8cm
\begin{center}
%\mbox{\epsfysize 4cm \epsffile{tqft.eps}} 
%RREMETtre tqft2.Eps
\mbox{\epsfysize 4cm \epsffile{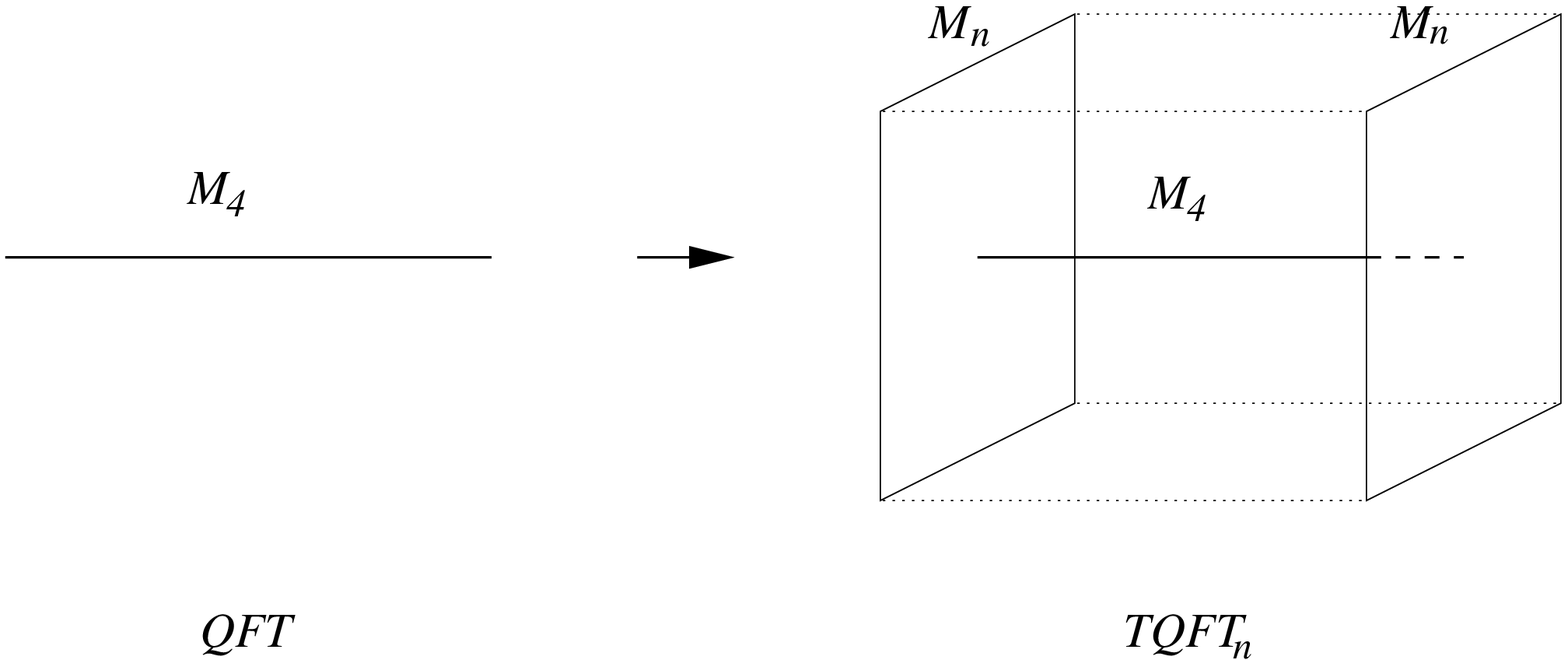}} 
\vskip .7cm\end{center}
 The interest of this generalization, when $n$ is larger
than one, is yet to be discovered, but we find that the  aesthetics of the
whole construction makes it worth being presented. 

\section{The fields}

We consider for simplicity the case of a commuting scalar field
$\phi (x)$   in 4~dimensions, with a Lagrangian $L_0(\phi)$ and action 
$I_0[\phi ]=\int d^4x L_0(\phi)$.
 We want to find another  formulation of  this theory with an action in a
space with dimensions $4+n$, where $n$ denotes generically the dimension
of the bulk, with observables that are defined in a slice of dimension 4.
The number of bosonic fields that are needed depends   on
$n$, and increases as $2^n$,  according~to:
\bea 
{\rm level} \ n=0, \quad \Phi_0 &=&     \phi_1(x) = \phi (x) \nonu  \\  
{\rm level} \ n=1, \quad  \Phi_1 &=&     \phi_1(x,t^1) ,\phi_2(x,t^1) 
\nonu\\ .... &\  &   \nonu\\
{\rm level} \ n,\ \ \ \ \ \quad  \Phi_n&=& \phi_1(x,t^1,\ldots,t^n   )
,\phi_2(x,t^1,\ldots,t^n),
\ldots \phi_{2^n}(x,t^1,\ldots,t^n)
\nonu\\
....&\  &  
\eea
Indeed, if we apply from level $n$ to $n+1$ the process explained
in~\cite{BaZw},    the number of field degrees of freedom  that are
needed doubles. Thus, we have a multiplet with $2^n$
components when the bulk has dimension $n$. For $n=1$,
  $\phi_2(x,t^1)$ can be heuristically interpreted as the Gaussian  noise of
a stochastic process. A more fundamental interpretation is that  
$\phi_2(x,t^1)$ is  the canonical moment with respect to the bulk time of
the  field  $\phi_1=\phi$, and so on.

At every level $n$, there is a hidden BRST symmetry, defined by the graded
differential operator $\Q$:
\bea
\matrix {
\Q\phi_p=\psi_p
& \ \ \ \ \ \ \ 
\Q\bar \Psi_{p+2^{n-1}} =\phi_{p+2^{n-1}}
\cr
\Q \Psi_p=0
&
\Q \phi_{p+2^{n-1}}  =0,
\cr
}
\eea
where, $1\leq p \leq 2^{n-1}$. The   $\Psi $'s and $\bar \Psi $'s are
ghosts and antighosts, with the opposite statistics to the $\phi$'s. The
set of fields
$\phi$, $\Psi$ and $\bar \Psi$ determine a BRST topological quartet. 
We can   define an anti-BRST operator, which merely  interchanges   the
ghosts and antighosts:
\def\bQ{ {\bar { s}_n}}
\bea
\matrix {
\ \  \bQ \phi_p=\bar \Psi_{p+2^{n-1} }
& \ \ \ \ \ \ \ \ 
\bQ  \Psi_{p } =-\phi_{p+2^{n-1}}
\cr
\bQ \bar \Psi_{p+2^{n-1}}=0
&\ \ \ 
\bQ \phi_{p+2^{n-1}}  =0
\cr
}
\eea
The action that describes the formulation of the
theory with a bulk of dimension $n$ must be $\Q$-exact, that is, it must
be of the form:
\bea
\int
d^4x dt^1\ldots dt^n \  \Q\big (\sum _{p=1}^{2^{n-1}}
\bar \Psi _{p+2^{n-1}} Z_p \big ),
\eea
that is,
\bea
\int
d^4x dt^1\ldots dt^n \Big(
 \sum _{p=1}^{2^{n-1}}
\phi _{p+2^{n-1}} Z_p
-
 \sum _{p=1}^{2^{n-1}}
\bar \Psi _{p+2^{n-1}}\Q Z_p  \Big )
\eea
To determine the action, it is sufficient to determine the expression of
the functionals
$ Z_p[\Phi_n]$, for every level $n$. At level $n$, the ghosts have
parabolic propagators along $t_n$. Thus, they can be integrated out exactly
 when one computes   correlations functions of  the
$\Phi$'s. Indeed, the latter cannot contain closed loops of the ghosts.
It follows that 
it is sufficient to  determine the following part of the action:
\bea
I_n=
\int
d^4x dt^1\ldots dt^n 
 \sum _{p=1}^{2^{n-1}}
\phi _{p+2^{n-1}} Z_p. 
\eea
The rest of the action  will be determined by the requirement of BRST symmetry.
The determination of the factors  $Z_p$ is  restricted by power counting and by 
symmetries. The latter include the  parity in the bulk, that is,  the
invariance under  
$t_p
\to -  t_p$, for
$1 \leq  p
\leq n$, and by  translation invariance. The covariance of the fields under
this parity will be shortly determined.

We define  the following self-consistent assignments for
the   dimensions of the bulk coordinates and of the fields: 
\bea\label{powerc}[t^n]^{-1}= 2^n\quad\quad\quad [\phi_p]=2p-1
\eea
With this assignments, the dimension of the Lagrangian at 
 level $n$  must be $2+2^{n+1}$.  This  ensures that the theory generated by
the action $I_n$ is renormalizable by power counting, as a generalization
 of~\cite{BaZw}. It is
of relevance to note that:
\bea
[\phi_1] +[\phi_{2^{n-1}}]+[t^n]^{-1}=  2+2^{n+1} 
\eea
This will imply that, in  the formulation at level $n$,  
$\phi_{2^{n-1}}$ is  the conjugate momentum of $\phi_1$ with respect
to $t^n$. This turns out to be one of  the key facts  for 
proving by induction that the physical content of the  theory at level $n$
is  the same
physics as for  the theory at level $n-1$, and so on, down to the ordinary
formulation with the action $I_0$.

\section{The action at  level $n$}

At level zero, the theory is defined by the standard  action
$I_0[\phi]=\int d^4x L_0(\phi(x))$. At level $n=1$, it is defined by:
\bea
I_1[\phi_1,\phi_2]=\int d^4xdt^1\ 
\Big  (
\phi_2\pa_1 \phi_1 +\phi_2\  (\phi_2+
 {{ \de I_0}\over {\de\phi_1}}\ )\ \Big )
\eea
The exponential of this action must be inserted in the path integral
with measure $[d\phi_{1 }]_{ x,t}[d\phi_{ 2}]_{x,t}$.
$I_1$   satisfies  power counting according to equation~(\ref{powerc})
(here
$[t^1]^{-1}=2$,
$[\phi_1]=1, [\phi_2]=3$) and   the 
bulk-parity symmetry $P_1$   is:
  \bea
 t^1 &\to& -t^1  \nonu
\\
\phi_1& \to& \phi_1 \nonu
\\
\phi_2 &\to& -\phi_2 - {{ \de I_0}\over {\de \phi_1}}
\eea
The action $I_1$ and its symmetry $P_1$  have been discussed
in details in~\cite{BaZw}, where we have also shown that it describes the
same  physics as the action $I_0$. Notice that the existence of the
symmetry
$P_1$ is obvious after the  elimination of $ \phi_2$ by its equation of
motion.

At level $n= 2$, the action is:
\bea\label{actiondeux}
I_2[\phi_1,\phi_2,\phi_3,\phi_4]=\int d^4xdt^1dt^2\ 
\Big(
\phi_3\pa_2 \phi_1 +\phi_3\   (\phi_3+
 {{ \de I_1}\over {\de \phi_1}}\ )\cr
+\phi_4\ ( \phi_2+{{ \de I_0}\over {\de \phi_1}}+\pa_1\phi_1
\ ) \Big)
\eea
$I_2$ is invariant under the bulk-parity   transformations $P_2$ and $P_1$.
  $P_2$ is defined as:
\bea
t_2   &\to& -t_2
\nonu  \\
\phi_1, \phi_2  &\to&   \phi_1, \phi_2 
\nonu  \\
\phi_3  &\to& - \phi_3-{{ \de I_1}\over {\de \phi_1}}
\nonu  \\
 \phi_4  &\to&   \phi_4  +\pa_2 \phi_2
\nonu  \\
\eea
The action of the  symmetry  $P_2$  on $\phi_4$  is such that 
 $\delta I_2=-\int \partial _2( I_1)$.
$P_1$ is
defined as:
\bea\label{paritedeux}
t_1   &\to& -t_1
\nonu  \\
\phi_1, \phi_3  &\to&   \phi_1, \phi_3 
\nonu  \\
\phi_2  &\to& - \phi_2-{{ \de I_0}\over {\de \phi_1}}
\nonu  \\
 \phi_4  &\to&   -\phi_4  +\phi_3 {{ \de^2 I_0}\over {\de \phi_1   \de \phi_1}}
\eea 
  $P_1$ transforms  $\phi_4$   in such a way that 
the variation of the  term $\phi_4( \delta  I_2/ \delta \phi _2)$ compensates 
that of  $\phi_3 ( \delta I_1 / \delta  \phi _1) $.

The parity  symmetry under $P_1$ and $P_2$ implies that a $I_2$ has the form displayed
in eq.~(\ref{actiondeux}).  In this action, power conting 
implies that  $I_0[\phi]$ is a local functional, which can be identified 
 as an  action that  is   renormalizable in 4 dimensions.
Thus, no  new parameter of physical relevance  can  be
introduced when one switches from the ordinary formulation to the
formulation with a bulk.  By a
straightforward generalization of~\cite{BaZw}, one can than  prove
that the correlations functions,  computed from $I_2$  at a given point of the
two-dimensional bulk,  satisfy the same Dyson--Schwinger equations as
those  computed from
$I_1$, at a given point of the one-dimensional bulk. In turn, there is 
 the  equivalence of the physics
computed either from 
$I_1$ from $I_0$, which gives the desired result that we can
use a two-dimensional bulk to compute physical quantities with the same
result as in the ordinary formulation. .

We can now give the general expression of the action at level $n$, which
satisfies power counting and is invariant under all parity transformations
in the  bulk, $t_p\to -t_p$, for $1\leq p\leq n$. It reads:
 \bea\label{actionN}
I_n=\int d^4xdt^1dt^2\ldots dt^n\ 
\Big(
&
\phi_{1+2^{n-1}}\pa_n  \phi_1 +\phi_{1+2^{n-1}}
\   (\phi_{1+2^{n-1}} + {{ \de I_{n-1}}\over {\de \phi_1}})\nonu\\
&+
\sum_{p=2+2^{n-1}} ^{2^n}
\phi_p
 {{ \de I_{n-1}[\phi_1 \ldots,\phi_{2^{n-1}}]}\over {\de \phi_{p-2^{n-1}}  
}}\  )
\Big)
\eea
$I_n$ is  invariant under the   parity transformation  $P_n$,  with:
\bea
t_n   &\to& -t_n
\nonu  \\
\phi_{1+2^{n-1}} &\to&  - \phi_{1+2^{n-1}} 
- {{ \de I_{n-1}}\over {\de \phi_1}}
\nonu  \\
\phi_p  &\to&   \phi_p  
\quad \quad  \  \hskip 2.05cm {\rm for} \    p <
1+2^{n-1}
\nonu  \\
\phi_p  &\to&   \phi_p +\pa_n \phi_{p-2^{n-1}}
\quad \quad   {\rm for}\  p >
1+2^{n-1}
\nonu  \\
\eea
Under the symmetry $P_n$, the Lagrangian density varies by a pure
derivative,
\def\L{{\cal L}}
${\L}_n\to {\L}_n -\pa_n({\L}_{n-1})$.

 As for the rest of the parity transformations $P_p$ of the fields in the
bulk, with $1\leq p\leq n-1$, their existence can be proven by induction.

Assume that the full bulk parity symmetry  exists at level $n-1$, that is,
field transformations exist that leave invariant 
$I_{n-1}[\phi_1,\ldots,\phi_{2^{n-1}}]$ for all transformations  $t_p\to
-t_p$, $1\leq p\leq n-1$. Then, the triangular nature of the Jacobian of
the transformation  $(\phi_1,\ldots,\phi_{2^{n-1}}) \to
P_p(\phi_1,\ldots,\phi_{2^{n-1}})$ at level $n-1$ implies that one can
extend this  transformation law for   the new fields that occur at level
$n$,
$(\phi_1,\ldots,\phi_{2^{n}}) \to P_p(\phi_{1},\ldots,\phi_{2^{n }})$  and that
$I_n$, as given in eq.~(\ref{actionN}), is invariant under $P_p$,    in a
way that generalizes ~eq.(\ref{paritedeux}).

 Conversely, the   parity symmetry  and  power counting imply that
$I_n$ must be  of the form~(\ref{actionN}). This shows that the number of
parameters of the theory  is  the same in bulk quantization, with any
given choice of the the bulk dimension n, as in the standard formulation.
 These
parameters are just those of an action that is renormalizable by power
counting in 4 dimensions.

We can now  write the action in the following form:
\bea\label{QactionN}
 \int d^4xdt^1dt^2\ldots dt^n\  \Q
\Big(
  &\bar \Psi_{1+2^{n-1}}
(\pa_n  \phi_1 +\phi_{1+2^{n-1}}
{{ \de I_{n-1}[\phi_1\ldots,\phi_{2^{n-1}}]}\over {\de \phi_{1}   }} )
\nonu \\
& +   
\sum_{p=2 } ^{2^{n-1}}
\bar\Psi_p
 {{ \de I_{n-1}[\phi_1\ldots,\phi_{2^{n-1}}]}\over {\de \phi_{p}   }}\   \
\Big) 
\eea

A propagation occurs in the new direction
$t^n$, while the equation of motions of the formulation at degree $n-1$ are
enforced in a BRST invariant way. Because the action is $\Q$ exact,
and    $\phi_{1+2^{n-1}}$ is the momentum with respect to $t_n$
of $\phi_1$,  the correlation functions that one can compute  in
the ($4+n$)-dimensional theory,   at an equal bulk component $t_n$, are
identical to those computed in the theory defined by   $I_{n-1}$. 
The proof is just as in the case of a one-dimensional bulk,
and uses  
the BRST invariance and    the   translation and parity symmetries in
the bulk. Finally,  the correlation
functions computed in the ($4+n$)-dimensional theory,   where all argument
only involve a single   point $T$ in the  bulk, are
identical to those computed from the basic four dimensional $I_0$, that is,
\bea
G _N^{I_0}\Big(x_1,\ldots,x_N\Big)
=
G _N^{I_n}\Big( (x_1,{  t^{p_1}})\ldots,(x_N, {  t^{p_N}})\Big)
{\Large {|}}_{{ t^{p_1}}=\ldots={  t^{p_N}}=T^{p}}
\eea
Due to translation invariance, such a correlation function  is 
independent on the
choice of  $T$.
 An another interesting expression of the action at level
$n$ is:
\bea\label{QactionNb}
 \int d^4xdt^1dt^2&\ldots& dt^n \ 
 (
 \Q   (  \bar \Psi_{1+2^{n-1}}
 \pa_n  \phi_1 
)    \cr &&
+ 
\Q\bQ   (   \bar \Psi_{1+2^{n-1}}\Psi_1 +
I_{n-1}[\phi_1,\ldots,\phi_{2^{n-1}}] )
\ )  
\eea

It shows that the Hamiltonian at level $n$ is a supersymmetric
term, $H~=~\demi\{Q_n,\bar Q_n\}$, which involves in the very simple way
the action at level
$n-1$.

%In this letter, we have shown that the quantization  with an additional
%time can be defined 

{\bf Acknowledgments}: This is a pleasure to thank Daniel Zwanziger 
for discussions related to this work. The research of Laurent Baulieu was
supported in part by DOE grant DE-FG02-96ER40959.

\end{document}